\documentclass[aps,prb,twocolumn,showpacs]{revtex4}%
\usepackage{amsfonts}
\usepackage{amsmath}
\usepackage{amssymb}
\usepackage{graphicx}%
\begin{document}
\title{Magneto-transport study of intra- and intergrain transitions in the magnetic
superconductors RuSr$_{2}$GdCu$_{2}$O$_{8}$ and RuSr$_{2}$(Gd$_{1.5}$%
Ce$_{0.5}$)Cu$_{2}$O$_{10}$}
\author{S. Garc\'{\i}a}
\affiliation{Laboratorio de Superconductividad, Facultad de F\'{\i}sica-IMRE, Universidad
de La Habana, San L\'{a}zaro y L, Ciudad de La Habana 10400, Cuba.}
\author{J.E. Musa}
\affiliation{Centro Brasileiro de Pesquisas F\'{\i}sicas, Rua Dr. Xavier Sigaud 150, Rio de
Janiero, RJ 22290-180, Brazil}
\author{R.S. Freitas and L. Ghivelder}
\affiliation{Instituto de F\'{\i}sica, Universidade Federal do Rio de Janeiro, C.P. 68528,
Rio de Janeiro , RJ 21941-972, Brazil}

\begin{abstract}
A characterization of the magnetic superconductors RuSr$_{2}$GdCu$_{2}$O$_{8}$
[Ru-(1212)] and RuSr$_{2}$Gd$_{1.5}$Ce$_{0.5}$Cu$_{2}$O$_{10}$ [Ru-(1222)]
through resistance measurements as a function of temperature and magnetic
field is presented. Two peaks in the derivative of the resistive curves are
identified as intra- and intergrain superconducting transitions. Strong
intragrain granularity effects are observed, and explained by considering the
antiphase boundaries between structural domains of coherently rotated
RuO$_{6}$ octahedra as intragrain Josephson-junctions. A different field
dependence of the intragrain transition temperature in these compounds was
found. For Ru-(1212) it remains unchanged up to 0.1 T, decreasing for higher
fields. In Ru-(1222) it smoothly diminishes with the increase in field even
for a value as low as 100 Oe. These results are interpreted as a consequence
of a spin-flop transition of the Ru moments. The large separation between the
RuO$_{2}$ layers in Ru-(1222) promotes a weak interlayer coupling, leading the
magnetic transition to occur at lower fields. The suppression rate of the
intragrain transition temperature is about five times higher for Ru-(1222), a
result we relate to an enhancement of the 2D character of the vortex
structure. A distinctive difference with conventional cuprates is the sharp
increase in amplitude of the intergrain peak in both systems, as the field is
raised, which is ascribed to percolation through a fraction of high quality
intergrain junctions.

\end{abstract}
\received[Received text]{26 March 2003}

\maketitle

\section{INTRODUCTION}

The ruthenate-cuprates systems RuSr$_{2}$RCu$_{2}$O$_{8}$ [Ru-(1212)] and
RuSr$_{2}$(R,Ce)$_{2}$Cu$_{2}$O$_{10}$ [Ru-(1222)], where R = Gd, Eu, are
currently receiving a great deal of attention. The onset of bulk
superconductivity in the presence of a ferromagnetic (FM)
component\cite{Felner97, Tallon} makes these compounds particularly suitable
to study the interplay between these usually exclusive phenomena. Ru-(1212) is
obtained from orthorhombic YBa$_{2}$Cu$_{3}$O$_{7-\delta}$ (YBCO) by full
replacement of Cu(1) sites at the chains for Ru ions, which add two oxygen
atoms to their neighborhoods in such a way that the original square
coordination of Cu(1) sites evolves to RuO$_{6}$ octahedra and the structure
becomes tetragonal.\cite{Cava} The Ru-(1222) structure is obtained from
Ru-(1212) by inserting a fluorite type (R$_{1-x}$Ce$_{x}$)$_{2}$ block instead
of the R plane.\cite{Cava} In both Ru compounds only one distinct Cu site with
fivefold pyramidal coordination exists [corresponding to the Cu(2) site in
YBCO]. The result is a sequence of magnetic RuO$_{2}$ planes between CuO$_{2}$
superconducting bilayers. Long-range order of the Ru magnetic moments occurs
at T$_{M}$ $\sim$133 K in Ru-(1212), followed by a resistive superconducting
(SC) transition at T$_{SC}$ $\sim$45 K.\cite{Bernhard} For Ru-(1222), T$_{M}$
ranges between 125 and 155 K, depending on the Ce content, while T$_{SC}$
moves around 30 - 35 K.\cite{Felner2002} Since in these compounds, for the
first time, T$_{M}$
$>$%
$>$
T$_{SC}$, it has been proposed that the superconducting transition leads
directly to the mixed state by spontaneous vortex phase (SVP) formation when
the internal magnetization exceeds the first critical field.\cite{Bernhard2,
Felner98}

The detection of a sizeable Meissner signal in Ru-(1212) strongly depends on
sample preparation conditions. When observed, it appears at about 15-30 K
below the resistive superconducting transition. SVP formation, large magnetic
penetration length and reduced effective grain size are claimed to account for
this behavior.\cite{Bernhard2} The latter is related to the existence of
intragrain domains, where the RuO$_{6}$ octahedra are coherently rotated
around the \textit{c}-axis, being separated by sharp antiphase
boundaries.\cite{McLaughlin} This characteristic is also observed in
Ru-(1222).\cite{Knee}

A major open issue is how the SC state is established in a grain in the
presence of a well-developed long-range magnetic order with a FM contribution.
Preliminary resistivity measurements for Ru-(1212) in zero external fields
suggest that a single grain behaves as a disordered Josephson-junction array
(JJA).\cite{Xue} Recent ac susceptibility experiments also support this
idea.\cite{Lorenz} In these reports, phase separation into antiferromagnetic
(AFM) and FM domains was suggested to be the source of such behavior. However,
this conjecture does not provide a consistent explanation of the
magneto-transport properties, as discussed below. Instead, we propose that the
Josephson-like behavior of the intragrain transition has a structural nature
in both ruthenate-cuprate systems, and that the results obtained from a
careful study of the resistive SC transition in the presence of a dc magnetic
field can be consistently explained with this approach.

Magnetotransport measurements are useful in order to understand the magnetic
structure in the ruthenate-cuprates, which is still a controversial subject,
by probing how the changes in the spin order of the Ru sub-lattice are
reflected in the superconducting transition. This motivated our investigation
of the magnetotransport of both ruthenate-cuprate systems. In this work we
present a systematic study of the resistive superconducting transition in
RuSr$_{2}$GdCu$_{2}$O$_{8}$and RuSr$_{2}$Gd$_{1.5}$Ce$_{0.5}$Cu$_{2}$O$_{10}$
in the presence of magnetic fields up to 9 T. In particular, the field
dependence of the position, amplitude and width of the peaks observed in the
derivative of the resistive curves is analyzed in detail. Also, ac magnetic
susceptibility measurements were performed in Ru-(1212) with the same
superimposed dc fields, looking for a correlation between transport and
magnetic properties. For Ru-(1222), it was not possible to establish a clear
correlation with ac magnetic susceptibility measurements, since this compound
exhibits unique dynamic features and metastable magnetic states which
complicate the interpretation.\cite{Ziv} Gd was chosen instead of Eu as the
rare earth element because the phase composition of the Eu-based compounds is
usually slightly poorer, exhibiting a small amount of ferromagnetic
SrRuO$_{3}$. The paramagnetic contribution of Gd does not affect the
conclusions. An YBCO sample was included in the study not only as a
conventional cuprate reference, but also because of its close structural
relationship with the ruthenate-cuprates. Intra- and intergrain transitions
were identified and their magnetic field dependence explained in terms of the
effects of the changes in the internal Ru magnetization on the antiphase
boundaries, which we propose to act as weak links between structural domains.
The presence of the fluorite type (Gd$_{1.5}$Ce$_{0.5}$) block in Ru-(1222),
which enhances the 2D character of both the vortex structure and the magnetic
order in the RuO$_{2}$ layers, is a relevant structural detail to explain the
differences with Ru-(1212).

\section{EXPERIMENTAL}

Polycrystalline samples of RuSr$_{2}$GdCu$_{2}$O$_{8}$ and RuSr$_{2}$%
Gd$_{1.5}$Ce$_{0.5}$Cu$_{2}$O$_{10}$ were prepared by conventional solid-state
reaction with high purity RuO$_{2}$, SrCO$_{3}$, Gd$_{2}$O$_{3}$, CeO$_{2}$
and CuO powders. The initial mixtures were decomposed at 960 $^{o}$C in air.
After milling and pressing operations, the material was reacted in flowing
nitrogen at 1000 $^{o}$C for 12 hours to avoid SrRuO$_{3}$ formation.
Sintering was performed at 1050 $^{o}$C for 4 days in flowing oxygen for
Ru-(1212) and at 1060 $^{o}$C for Ru-(1222), followed by cooling at a rate of
45 $^{o}$C/ hour. All the samples show a density higher than 70 \% the
crystallographic one.

Room temperature x-ray diffraction patterns were collected to check phase
composition in a Rigaku powder diffractometer in step-scanning mode
($20^{o}\leq2\theta\leq80^{o}$). The microstructure of the samples was probed
using scanning electron microscopy (SEM). Resistance and ac susceptibility
measurements were performed in a Quantum Design PPMS system, with the
following dc magnetic field values: H = 0. 0.01, 0.03, 0.1, 0.3, 1, 3, 6 and 9
T, with an ac amplitude h$_{ac}$ = 0.1 Oe. The resistance was measured using a
standard four-probe technique, with a polarization current of 0.1 mA. No
temperature hysteresis effects were observed. The derivative of the resistive
curves was obtained by conventional numerical calculations.\cite{Pureur}

\section{RESULTS}

The room temperature x-ray diffraction patterns correspond to YBa$_{2}$%
Cu$_{3}$O$_{7-\delta}$, RuSr$_{2}$GdCu$_{2}$O$_{8}$ and RuSr$_{2}$Gd$_{1.5}%
$Ce$_{0.5}$Cu$_{2}$O$_{10}$, with no spurious lines being observed. The SEM
images are shown in Fig. 1. For YBCO (top panel), a quite dense arrangement of
parallelepiped-shaped grains with sharp edges ($\sim$5x10x10 $\mu$m$^{3}$ in
size) was observed. The microstructure of Ru-(1212) shows a relatively uniform
size distribution with rounded grains of about 1-3 $\mu$m (middle panel). Some
grains form compact agglomerates, which are well connected, leaving clearly
distinguishable intergranular regions. For the Ru-(1222) sample (bottom
panel), the microstructure exhibits a more dense packing due to the presence
of a significant fraction of crystallites of $\sim$0.5-1 $\mu$m in size
surrounding larger grains and filling the space between them. The larger
grains are similar in average size to those observed for Ru-(1212). The small
crystallites do not correspond to a secondary phase, since they are present in
such extent that would become detectable in the x-ray diffraction measurements.

Figure 2 shows the temperature dependence of the resistance R(T, H) in the
region of the superconducting transition for YBCO (a), Ru-(1212) (b), and
Ru-(1222) (c); the values of the applied dc magnetic field are indicated.
Their respective derivative curves, at constant field, are shown in Fig. 3
(note a more fine temperature scale for YBCO). Several features are relevant
in the derivative curves: the number of peaks and their widths for H = 0 and
the field dependence of their amplitudes, widths and positions. We first
consider the reference YBCO sample. In this case, there is a sharp peak at
T$_{1}$ = 91 K for H = 0, with a full width-half maximum $\Gamma$ $\sim$0.3 K
[see Fig. 3(a)]. On cooling, this peak is followed by a much broader and less
intense one, located at T$_{2}$ = 88.5 K. Both T$_{1}$ and T$_{2}$ are
indicated by arrows in the inset of Fig. 3(a). Traditionally, the existence of
two peaks has been interpreted as intra- and intergrain
SC-transitions.\cite{Early} In the following, we denote the high (T$_{1}$) and
low (T$_{2}$) temperature maxima as intragrain and intergrain peaks,
respectively. With the increase in H up to 0.3 T, peak 1 slightly moves to
lower temperatures at a rate of $\sim$0.8 K/T, with a small reduction in
amplitude and little increase in $\Gamma$ ($\Gamma$ $\sim$0.7 K for H = 0.3
T). For H $\geqslant$1 T, shift and broadening become more evident, but the
magnitude of these effects is still $\sim$1-2 K; also, a further reduction in
amplitude is observed. The field dependence of the shift in T$_{1}$ [$\Delta
$T$_{1}$(H) = T$_{1}$(H) - T$_{1}$(0)] is shown in Fig. 4 for low fields (a)
and for the whole field range (b). T$_{2}$ continuously diminishes with the
increase in H with an initial slope of $\sim$20 K/T; Fig. 5 shows the $\Delta
$T$_{2}$(H) = T$_{2}$(H) - T$_{2}$(0) behavior. Finally, the interval between
the thermodynamic transition temperature T$_{th}$ and T$_{1}$ at zero field is
$\Delta$T$_{th,1}$ = T$_{th}$ - T$_{1}$(0) $\sim$1 K. T$_{th}$ is taken as the
value at which the derivative curve departures from the high temperature
baseline, yielding T$_{th}$\ =\ 93 $\pm$\ 0.25 K. Although T$_{th}$ is not
sharply defined due to the smooth shape of the derivative curve, there is no
doubt that $\Delta$T$_{th,1}$ is not greater than $\sim$\ 2 K.

For Ru-(1212) two broad overlapped peaks in the H = 0 curve were observed at
temperatures T$_{1}$ and T$_{2}$, as identified by arrows in Fig. 3(b), where
T$_{th}$ ($\simeq$ 55 $\pm$\ 1 K) is also indicated. At first sight, this
feature is absent in the Ru-(1222) sample [see Fig. 3(c)], for which an
apparently single broad asymmetric peak is observed. However, the evolution of
the curves with the increase in H clearly reveals the presence of two peaks.
The estimation of $\Delta$ T$_{th,1}$ for Ru-(1212) is $\sim$12 K and $\sim$17
K for Ru-(1222) (T$_{th}$ $\simeq$ 45 $\pm$\ 1 K ), quite large in comparison
to YBCO. As reported from heat capacity measurements,\cite{Tallon2} T$_{th}$
does not diminish even for an applied field of 9 T.

The field dependence of the intragrain transition temperature T$_{1}$ was
found to be different in the Ru-based compounds. For Ru-(1212), the increase
in H up to 0.1 T leaves peak 1 unchanged in position, width and amplitude. For
H $\geqslant$ 0.3 T, T$_{1}$ smoothly diminish with an initial slope of $\sim
$7 K/T (one order of magnitude higher than for YBCO), while the peak strongly
broadens [Fig. 3(b)]. For Ru-(1222), T$_{1}$ diminishes as the field is
increased even for a value as low as 100 Oe, with an initial slope of $\sim$35
K/T. These features are clearly evidenced in Fig. 4(a) [low field range]. For
H
$>$
0.1 -- 0.3 T, the suppression rate of T$_{1}$ diminishes in the
ruthenate-cuprates, as can be seen in Fig. 4(b) [full range of field].

The intergrain transition temperature T$_{2}$ in the Ru-based systems rapidly
decreases for increasing fields up to H $\simeq$ 0.1 - 0.3 T, with a large
initial slope of $\sim$180 K/T. For H
$>$
0.3 T it diminishes at a much lower rate ($\sim$0.5 K/T), as shown in Fig. 5.
A relevant difference between the ruthenate-cuprate samples and YBCO is the
steep increase in amplitude of peak 2, accompanied by narrowing, for H
$>$
0.3 T [see Figs. 3 (b) and (c)].

Figure 6(a) shows the temperature dependence of the ac magnetic susceptibility
$\chi$(T,H) for Ru-(1212), with superimposed dc magnetic fields of the same
magnitude for which the resistive curves were measured. A diamagnetic
transition is very well defined for all H values at onset temperatures
T$_{o\chi}$, which match well with the zero resistance temperatures, as
determined from the resistive curves. Figure 6(b) shows an enlarged section of
the region of the superconducting transition. As T$_{o\chi}$(H) is approached
on cooling, the curves for H $\leqslant$ 0.1 T exhibit an upward deviation,
which magnitude continuously increases with the increase in field. For H
$\geqslant$ 0.3 T, the baseline of the curves is shifted to lower values,
while the deviation continuously evolves to a smeared drop at high fields. The
intragrain transitions temperatures, as determined from the peaks in the
derivative of the resistive curves, are indicated by arrows for H = 6 and 9 T.

\section{DISCUSSION}

The observed suppression rates for T$_{1}$ in both Ru-based compounds are
indicative that the intragrain superconductivity is due to a phase-lock
transition of a nanoscale JJA. Additionally, we note that the shape of the
T$_{1}$(H) curves in Fig. 4(b) are quite different from that expected for a
bulk superconductor, i.e., from Guinzburg-Landau theory. Phase separation into
nanoscale AFM and FM domains has been proposed\cite{Lorenz} as a possible
scenario for the Josephson-like behavior. In particular, the peak of positive
magnetoresistance observed for Ru-(1212)\cite{McCrone, Chen} is qualitatively
explained under this assumption. However, the absence of such a peak for
Ru-(1222), a system for which magnetic phase separation has also been proposed
to interpret thermal-magnetic memory effects,\cite{Xue2} indicates that this
interpretation is questionable. It should also be mentioned that $\mu$SR
experiments,\cite{Bernhard} which show that the internal magnetic field in the
compounds is uniform, provide strong indication against the existence of
magnetic domain segregation.

Alternatively, the Josephson-like behavior of the intragrain transition might
be explained in terms of a phase-lock process that occurs between structural
domains. As already mentioned, there are domains of coherently rotated
RuO$_{6}$ octahedra $\sim$14%
${{}^o}$
around the \textit{c}-axis, separated by sharp antiphase boundaries with local
distortions in \textit{both} Ru-(1212) (Ref. 8) and Ru-(1222)
systems.\cite{Knee} It has been shown that these structural domains are
relevant to explain the shift to lower temperatures of the Meissner drop in
Ru-(1212) in comparison to the resistive SC transition,\cite{Bernhard2} as a
consequence of SVP formation followed by flux expulsion from the structural
domains. Also, the temperature dependence of the microwave resistance for
Ru-(1212) in the region of the SC transition has been consistently interpreted
in terms of SVP formation.\cite{Pozec} For Ru-(1222), this mechanism has been
proposed to explain the dependence of the magnetic and transport properties on
the Ce concentration.\cite{Williams} Thus, we believe that there is strong
evidence to support the use of this approach to interpret our
magneto-transport measurements.

The measured single-phase x-ray diffraction patterns and the SEM results allow
us to rule out impurities or inhomogeneities effects as a possible cause for
the the large interval between the thermodynamic and the intragrain transition
temperatures, $\Delta$T$_{th,1}$, in zero external field, and therefore the
results presented are essentially determined by intrinsic properties of the
compounds. We interpret the large $\Delta$T$_{th,1}$ values for \textit{both}
Ru-based compounds in zero external field in terms of SVP formation induced by
the magnetization of the Ru-sub-lattice, followed by flux expulsion in the
domains. For temperatures below and near T$_{SC}$, the vortex lines created by
the internal field of the Ru sub-lattice are weakly pinned, and the Lorentz
force associated to the measuring current will drive a given fraction of them.
On cooling, pinning increases in the intragrain domains and flux lines are
gradually trapped. Also, the first critical field of the domains increases;
when it becomes higher than the internal magnetization, Meissner effect occurs
in the domains, with partial expulsion of the vortex lines, leading to flux
compression at the antiphase boundaries. The value of the local field at the
boundaries depends on the size of the neighboring domains and the amount of
flux trapped. The result is a complex thread of magnetic field lines across
the intragrain network, generating a variety of local effective fields. If the
boundaries act as Josephson junctions, the domains will gradually become
phase-locked as the temperature is decreased, until a maximum rate of the
percolation process is reached at T$_{1}$. The higher $\Delta$T$_{th,1}$ by
about 5 K for Ru-(1222) is attributed to the larger distance between the
CuO$_{2}$ planes. This structural feature enhances the 2D character of the
vortex lattice, promoting a less pinned structure. Lower temperatures will be
required to prevent dissipation associated to flux motion and to achieve a
stationary flux distribution at the interdomain boundaries, through which the
intragrain percolation may proceed. These effects in zero external field are
absent for YBCO, since there is neither SVP formation nor a domain structure.

\subsection{Intragrain transition in Ru-(1212)}

In the scheme depicted above the magnetization of the Ru sub-lattice is
essential in determining the details of the spontaneous vortex structure. The
fact that T$_{1}$ remains unchanged for Ru-(1212) up to H = 0.1 T suggests
that a) an external field of this strength has a little effect on the
magnetization of the RuO$_{2}$ layers at temperatures around 40-50 K ($\sim$90
K below T$_{M}$), and b) the effective internal field at the boundaries is
considerably higher than the applied magnetic field. At first sight, point b)
seems to be in contradiction with the fact that internal fields of only $\sim
$700 Oe have been measured at the Gd site by electronic paramagnetic resonance
(EPR) \cite{Fainstein} and at the so called apical site of the structure by
$\mu$SR measurements.\cite{Bernhard} However, when the Meissner effect is
established in the structural domains, a number of vortex lines are expulsed
from them and compressed into the thin thickness of the antiphase boundaries,
leading to a high local field. This is the actual value of field through which
a coherent SC state has to nucleate between domains. External fields H
$\lesssim$ 0.1 T make a negligible contribution to the interdomain field, and
no shift in the intragrain peak is observed.

The effect of an external field on the intragrain transition is important for
promoting a re-arrangement in the magnetic order of the Ru moments. The fact
that the $\chi$(T,H) curves change their behavior just at H $\simeq$ 0.1 - 0.3
T, supports this idea. We recall that for Ru-(1212) when a magnetic field H =
0.4 T is applied a change in the neutron diffraction pattern is observed, a
result interpreted as due to a spin-flop transition.\cite{Lynn} Also, detailed
magnetic measurements in this compound indicate that a spin-flop transition
should occur at a critical field of $\sim$0.14 T for crystallites with the
RuO$_{2}$ layers oriented parallel to the applied field.\cite{Butera} These
values are near to the fields at which we observed the decrease in T$_{1}$ and
the changes in the $\chi$(T, H) curves.

The features observed in the $\chi$(T, H) curves can also be explained in
terms of a spin transition. The evolution from an upward deviation to a drop
as the field is increased beyond 0.1 T implies that the relative contribution
to the net magnetization from components of different sign changes. The main
two contributions to the positive background are the FM component of the Ru
sub-lattice and the paramagnetic signal of the Gd moments. The fact that the
diamagnetic contribution becomes gradually detectable for H $\gtrsim$ 0.1 T
suggests that the FM component is reduced as a consequence of the spin
re-orientation, possibly including a change of the Ru moments from
\textit{c}-axis alignment to planar, as proposed from neutron diffraction
measurements.\cite{Lynn} According to this assumption, the magnetization at
the CuO$_{2}$ superconducting planes will be lowered, leading either to a
mixed state with a lower density of vortex lines, i.e., with an increased
fraction of the superconducting volume, or preventing SVP formation if the
internal magnetization becomes lower than the first critical field of the
domains. The boundaries will then be under the action of higher flux
compression, as it is expulsed in a larger extent. If the intragrain
transition is considered to be a phase-lock transition between structural
domains, lower temperatures will be required to achieve intragrain percolation
through a network of boundaries with an increased average local field. Also,
under this approach, the number of screened Gd paramagnetic ions will be
increased. Thus, the picture of a magnetic transition leading to a state with
a reduced Ru magnetization diminishes the positive components to the net
magnetization, increases the superconducting fraction in the domains, and
depletes the intragrain transition temperature. It is worth mentioning that
this analysis is independent of the actual order of the Ru moments below the
spin-flop transition, and additional microscopic results are needed to
interpret our magnetotransport measurements in terms of a well defined
magnetic structure.

\subsection{Intragrain transition in Ru-(1222)}

The decrease of T$_{1}$ in Ru-(1222) for a field as low as 100 Oe, as shown in
Fig. 4(a), suggests a different magnetic response of the Ru sub-lattice in
this compound. We believe that the enhancement of the 2D character of the
magnetic order of the RuO$_{2}$ layers due to the insertion of the (Gd,
Ce)$_{2}$O$_{2}$ fluorite block instead of the Gd plane in Ru-(1212), is a key
point to understand this behavior. The larger separation between the magnetic
layers in Ru-(1222) leads to a weak superexchange coupling between the
layers.\ In addition, differently than the case of Ru-(1212), any possible
chains would be also affected by the fact that the nearest-neighbor Ru ions
are not vertically aligned due to a shift induced by the fluorite block. Thus,
other mechanisms are needed to attain a long-range order of the Ru moments
along the \textit{c}-axis. Recently, an interlayer coupling via dipole-dipole
interaction has been proposed to explain the hysteretic behavior of Ru-(1222),
leading to a spin-flop of the Ru moments for magnetic fields in the 0-100 Oe
range,\cite{Ziv} a result which agrees with our interpretation. In addition,
magnetic frustration effects and spin-glass behavior have been claimed to
explain magnetic relaxation results in Ru-(1222), consistently with a weak
magnetic coupling in this system.\cite{Cardoso}

There is yet another important difference between the Ru-(1212) and Ru-(1222)
systems in relation to the magnetism of the RuO$_{2}$ layers which favors
T$_{1}$ suppression at low fields in the latter compound. For Ru-(1222), XANES
measurements\cite{Felner99} reveal the absence of a Ru$^{4+}$/Ru$^{5+}$ mixed
valence state, as observed in Ru-(1212).\cite{Liu} This precludes the
emergence of ferrimagnetic order in Ru-(1222). Ferrimagnetism in Ru-(1212) has
been claimed to be the source of the high magnetization measured for this
compound,\cite{Butera} which can not be explained only in terms of spin
canting considerations. These results are relevant to the present study
because they point to a state of low internal magnetization in Ru-(1222) at
zero external field. Unfortunately, the long-range order of the Ru moments in
Ru-(1222) is unknown; although neutron powder diffraction results were
reported,\cite{Knee} the exact magnetic structure has not been unveiled.

The approximately five times larger suppression rate of T$_{1}$ ($\sim$35 K/T)
at low fields in comparison to Ru-(1212) can be understood upon the same
considerations used to explain the larger $\Delta$T$_{th,1}$ interval in
Ru-(1222). As the external field is increased, the vortex structure in
Ru-(1222), which has a higher 2D character, will be depinned more easily.
Further cooling in comparison to Ru-(1212) would be required to attain a
stable configuration of flux lines across the interdomain boundaries, shifting
the intragrain transition to lower temperatures.

\subsection{Intergrain transitions}

Both compounds exhibit a rapid decrease of the intergrain transition
temperature T$_{2}$, for H $\lesssim$ 0.1- 0.3 T, with an initial slope which
is one order of magnitude higher than for YBCO, followed by a much smaller
suppression rate at higher fields (see Fig. 5). For Ru-(1212) it is possible
to establish a clear correlation between the T$_{2}$(H) behavior and the field
dependence of T$_{1}$. Since there are no changes in the intragrain transition
up to H = 0.1 T, the contribution of the Ru magnetization in the grains to an
effective field at the intergrain links remains essentially the same. Thus,
the decrease in T$_{2}$ in this interval is only due to the increase in the
external field. For H
$>$
0.1 T, the magnetic transition towards the proposed state of lower internal
magnetization gradually takes place in the polycrystalline sample with the
increase in field. This reduces the contribution of a fraction of the grains
to the local field at their neighboring intergrain junctions. The net effect
of an increasing external field acting on an intergrain network which improves
its connectivity as the intragrain magnetic transition proceeds is a lower
suppression rate in T$_{2}$. For Ru-(1222) the interpretation is less clear,
since in this case there is not a well defined magnetic field value at which
the re-arrangement of the Ru moments occurs, but instead a smooth decrease of
T$_{1}$. However, whatever the exact order of the Ru moments at low fields in
this system might be, the T$_{1}$(H) curve also shows a rapid variation in the
H $\simeq$ 0.1- 0.3 T interval, followed by a lower suppression rate [see Fig. 4(b)].

The field dependence of the amplitude of the intergrain peak in the Ru-based
compounds has a distinctive difference as compared with YBCO. Although it
regularly diminishes with the increase in field up to H = 0.1 T, accompanied
with broadening, as in the conventional cuprates, a steep increase and
narrowing are observed for H
$>$
0.3 T. In YBCO, the intergrain transition occurs over a wide distribution of
link qualities, which become gradually phase-locked on cooling, leading to a
broad peak. One possible explanation for the ruthenate-cuprates behavior in
the high field range is that low quality links would be inactivated
definitively due to the contribution of the magnetization in the grains to the
effective field at the junctions. In this scheme, the intergrain transition
would take place only through a fraction of high quality connections. This
fraction of \textquotedblleft available\textquotedblright\ links diminishes as
the external field is increased, and so the range of temperatures in which
percolation proceeds, leading to sharper transitions.

Another interesting feature of the field dependence of the intergrain
transition is that the T$_{2}$(H) curves, including YBCO, have a very similar
slope for H
$>$
0.3 T, as can be seen in Fig. 5, suggesting that the curves are displaced due
to the contribution of the magnetization in the grains to the effective field
at the intergrain links. The fact that for Ru-(1222) the reduction of T$_{2}$
for a given field is smaller than for Ru-(1212) is consistent with our
previous considerations about a lower magnetization in the grains for the
former compound. Additional studies are needed for a better understanding of
the role of the grain magnetization in the intergrain transition.

\section{CONCLUSIONS}

We presented data indicating that the intragrain transition in both Ru-(1212)
and Ru-(1222) exhibits a phase-lock behavior of a nanoscale Josephson-junction
array. It is proposed that the presence of domains of coherently rotated
RuO$_{6}$ octahedra, a common structural feature of both compounds, is the
source of such behavior. The differences in the field dependence of the
intragrain transitions in these systems are interpret in terms of the magnetic
response of the RuO$_{2}$ layers. The addition of the (Gd, Ce)$_{2}$ block in
Ru-(1222) is a key point to explain why a spin-flop transition occurs in this
material at a field one order of magnitude lower than for Ru-(1212), due to
the enhancement of the 2D character of the magnetic order in the RuO$_{2}$
layers. We argue also that this structural difference promotes a less pinned
vortex lattice, leading to a five times larger suppression rate in the
intragrain transition at low fields. The field dependence of the intergrain
transition temperature is consistent with the changes in the intragrain
magnetization in both compounds. The sharp intergrain peak at high fields
suggests that intergrain percolation occurs only through a fraction of high
quality junctions.\bigskip

\begin{acknowledgments}
S. Garc\'{\i}a acknowledges financial support from FAPERJ, CAPES-Brazil, and
CLAF (Latin American Centre of Physics). R.S. Freitas was supported by FAPERJ.
We thank A. L\'{o}pez and R. Cobas for help in sample preparation, and E.M.
Baggio-Saitovitch for fruitful discussions.
\end{acknowledgments}

\section{Figure Captions}

Figure. 1. Scanning electron microscopy images of YBa$_{2}$Cu$_{3}$O$_{7}$
(YBCO - top panel); RuSr$_{2}$GdCu$_{2}$O$_{8}$ (Ru-[1212] - middle panel);
and RuSr$_{2}$Gd$_{1.5}$Ce$_{0.5}$Cu$_{2}$O$_{10}$ (Ru-[1222] - bottom panel).

Fig. 2. Temperature dependence of the resistance, measured with dc magnetic
fields H = 0, 0.01, 0.03, 0.1, 0.3, 1, 3, 6, and 9 T: (a) YBCO, (b) Ru-(1212)
and (c) Ru-(1222).

Fig. 3. Derivative dR/dT of the resistive curves shown in Fig. 1: (a) YBCO,
(b) Ru-(1212) and (c) Ru-(1222). Inset in (a): an enlarged region of the
derivative curve for YBCO at zero external field. The intragrain (T$_{1}$) and
intergrain (T$_{2}$) peaks are identified in the inset and in panel (b) for H
= 0 T. The thermodynamic transition temperature T$_{th}$ is indicated for
Ru-(1212). The lines are guides to the eyes.

Fig. 4. Field dependence of the shift in the intragrain transition
temperature, $\Delta$T$_{1}$(H) = T$_{1}$(H) - T$_{1}$(0), as determined from
the derivative of the resistive curves of the studied samples: (a) an enlarged
section of the low field interval, and (b) for the whole range of fields. The
lines are guides to the eyes.

Fig. 5. Field dependence of the shift in the intergrain transition
temperature, $\Delta$T$_{2}$(H) = T$_{2}$(H) - T$_{2}$(0), as determined from
the derivative of the resistive curves. The lines are guides to the eyes.

Fig. 6 (a) Temperature dependence of the ac magnetic susceptibility for
Ru-(1212). The dc magnetic fields are the same used in the resistance
measurements; (b) an enlarged section of the region of the superconducting
transition. For H = 6 and 9 T the temperatures at which the corresponding
intragrain peaks occur in the derivative of the resistive curves are indicated
by arrows. The lines are guides to the eyes.

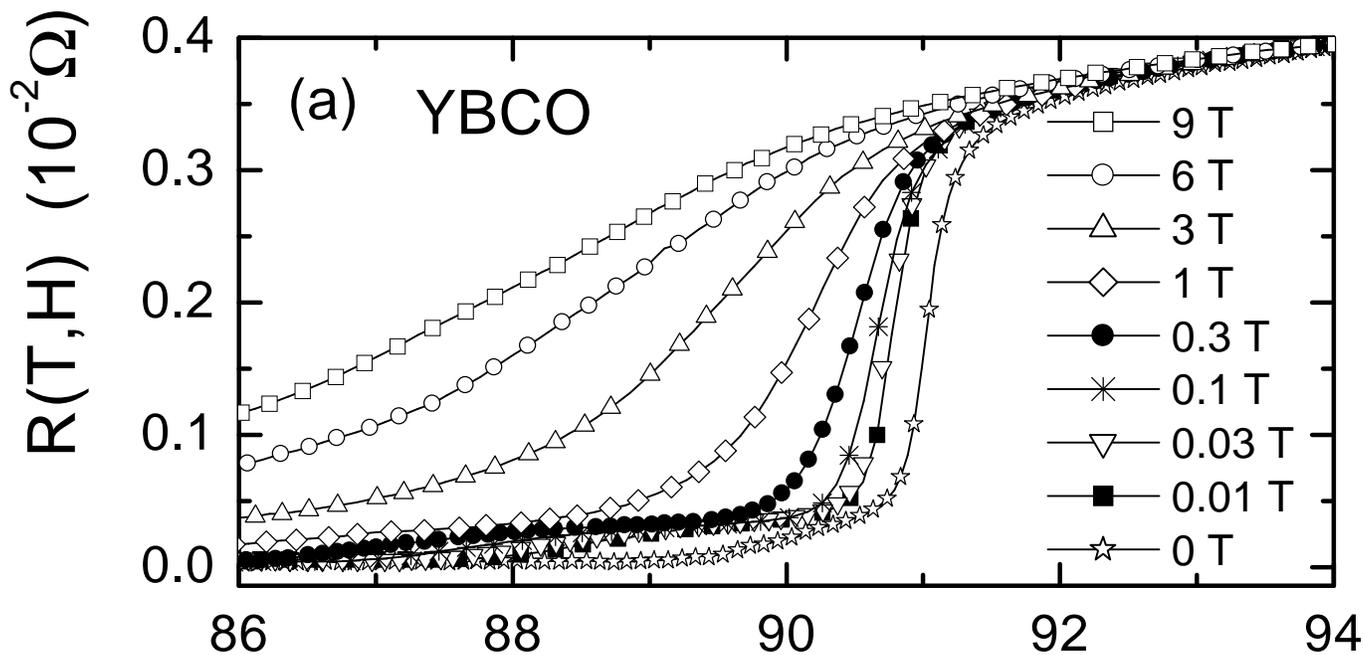
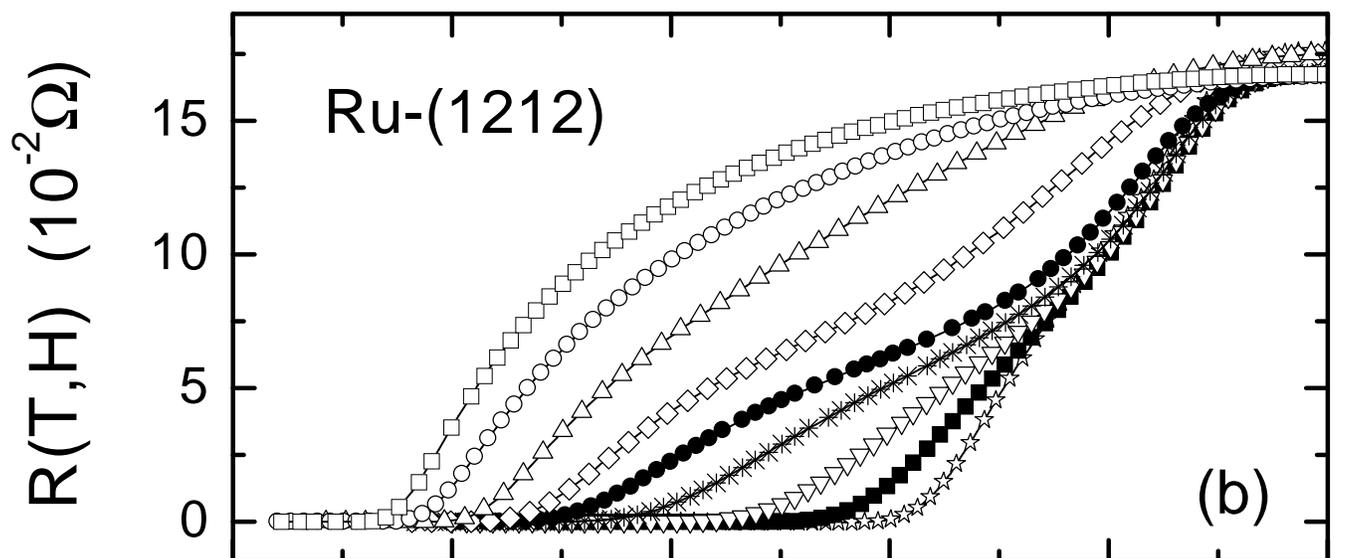
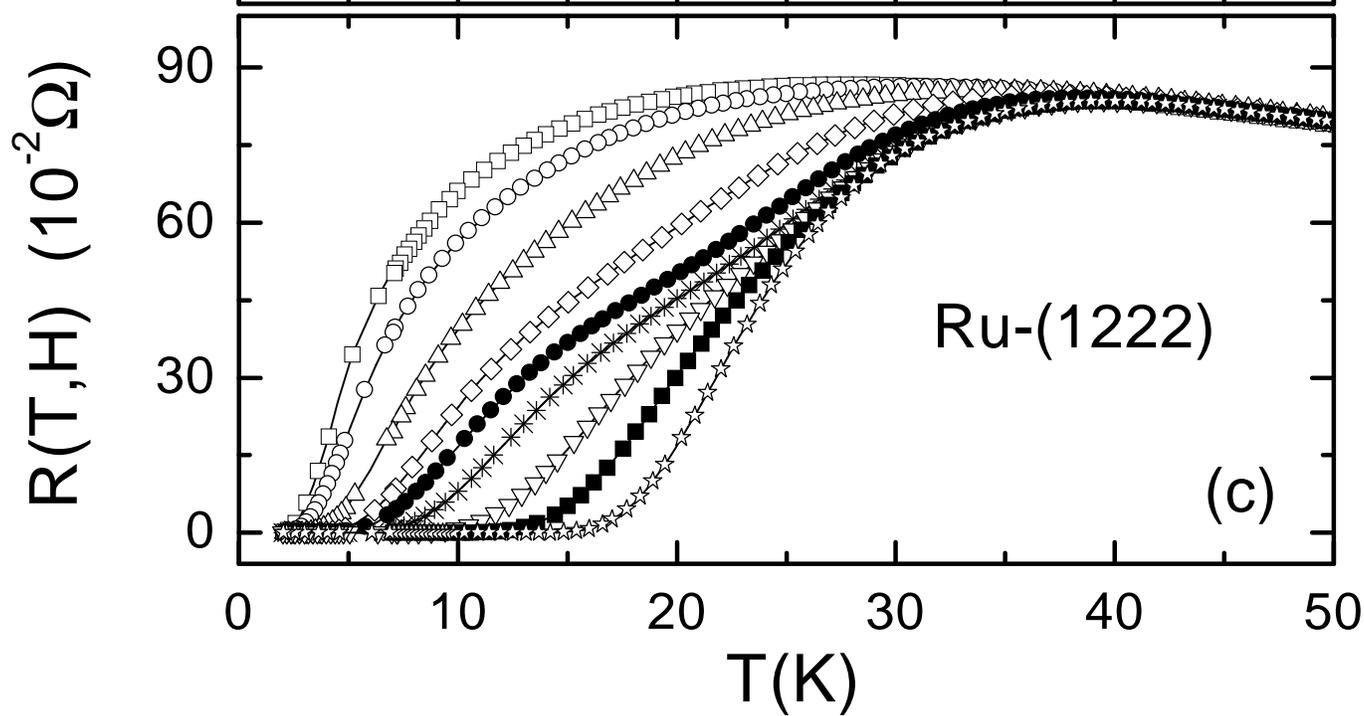
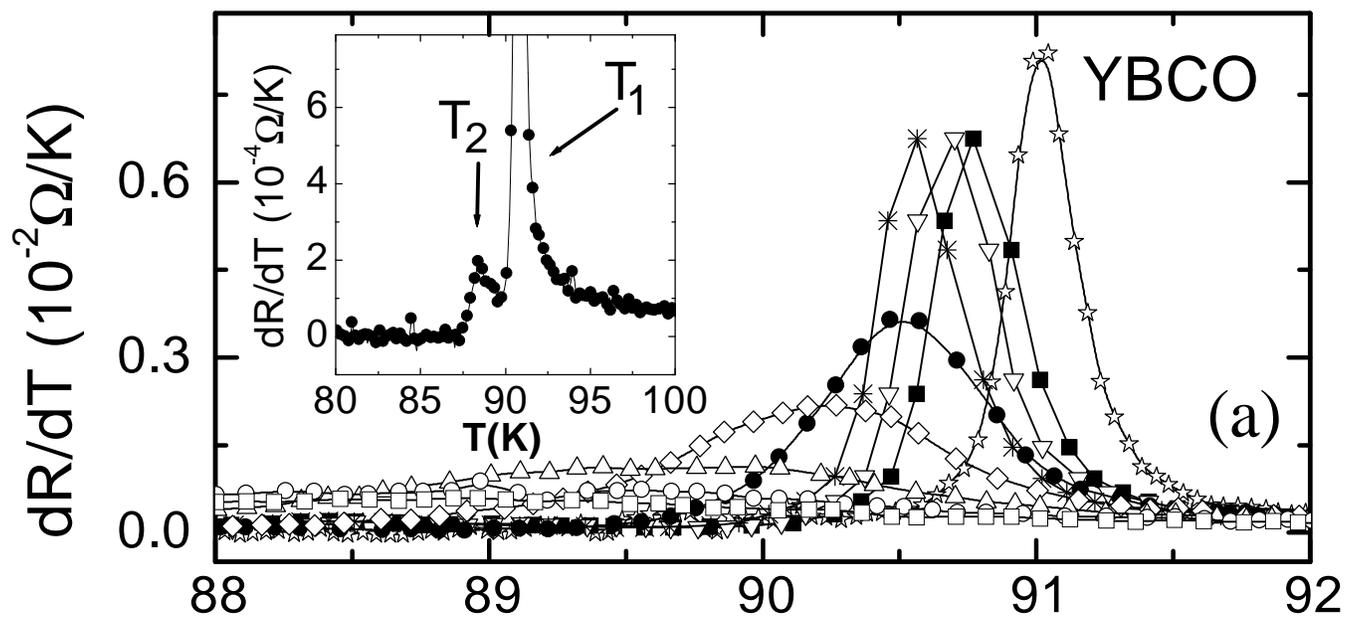
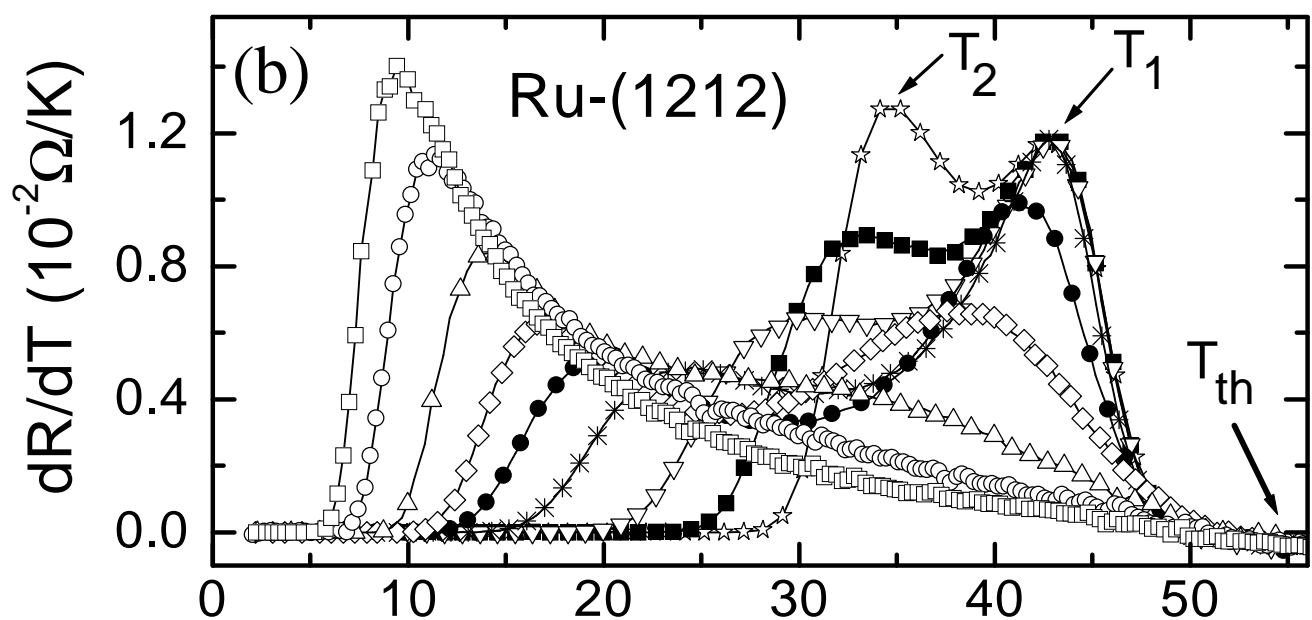
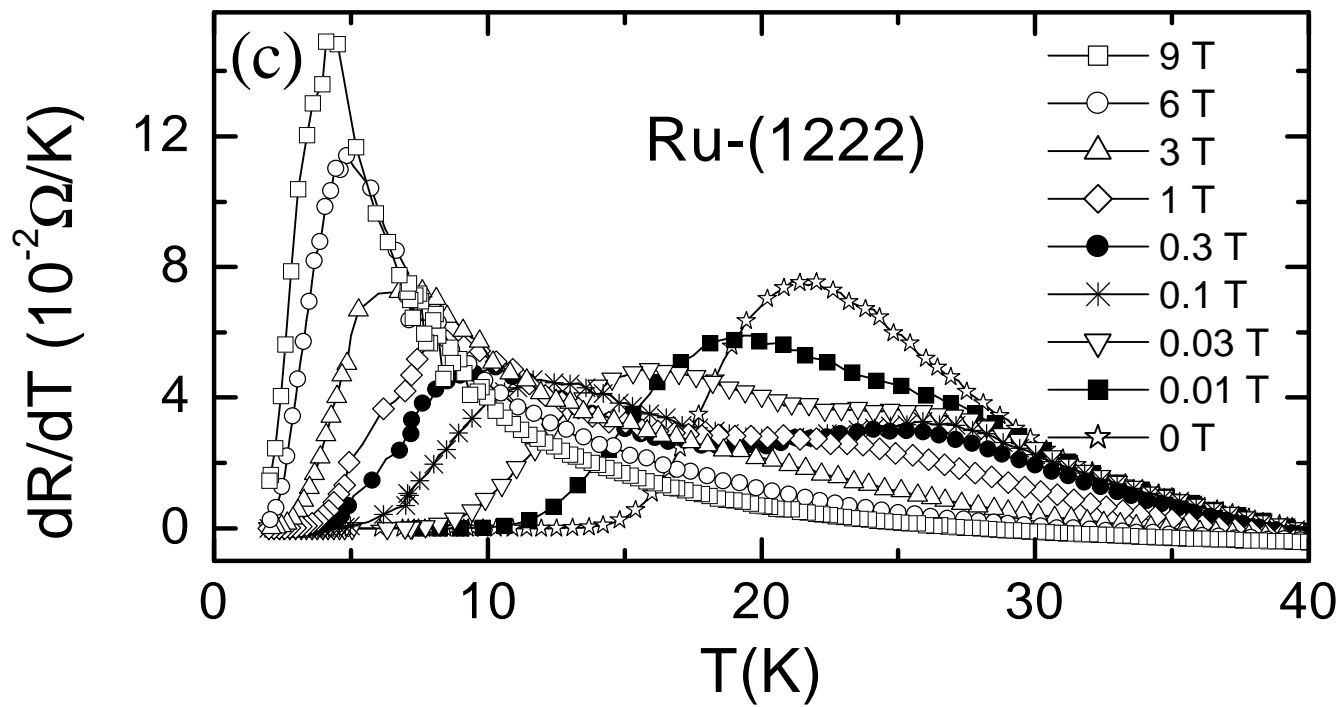
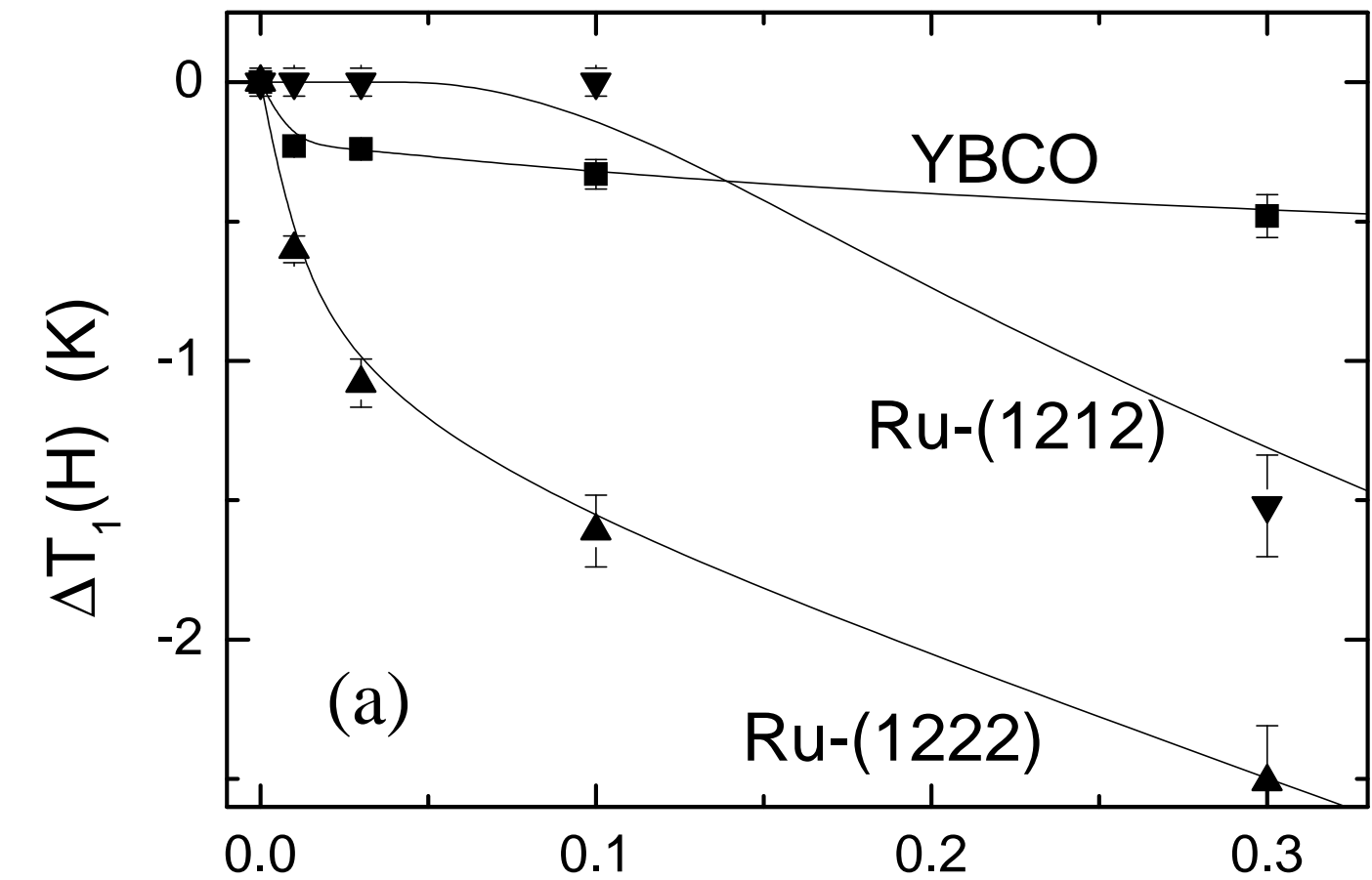
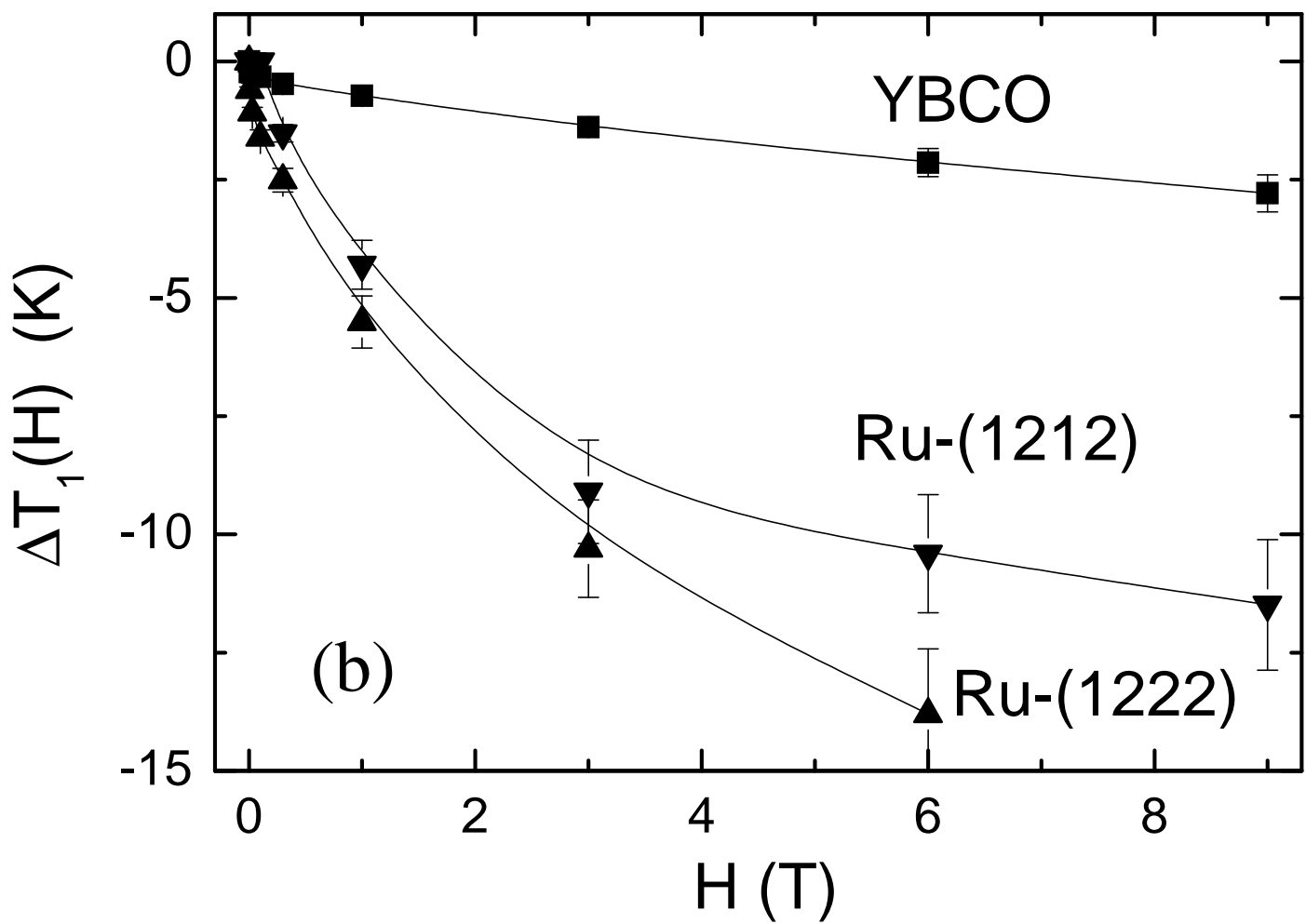
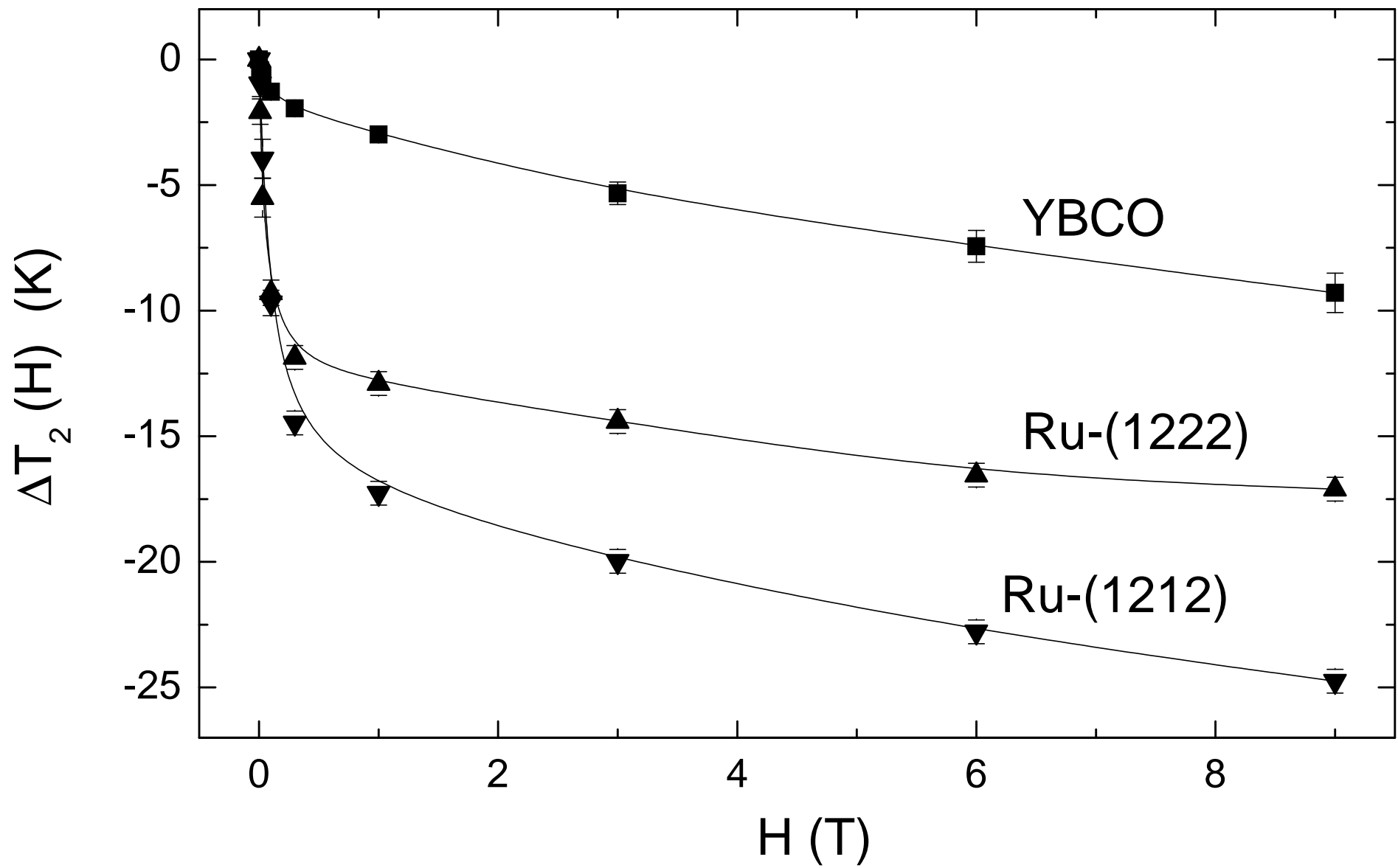
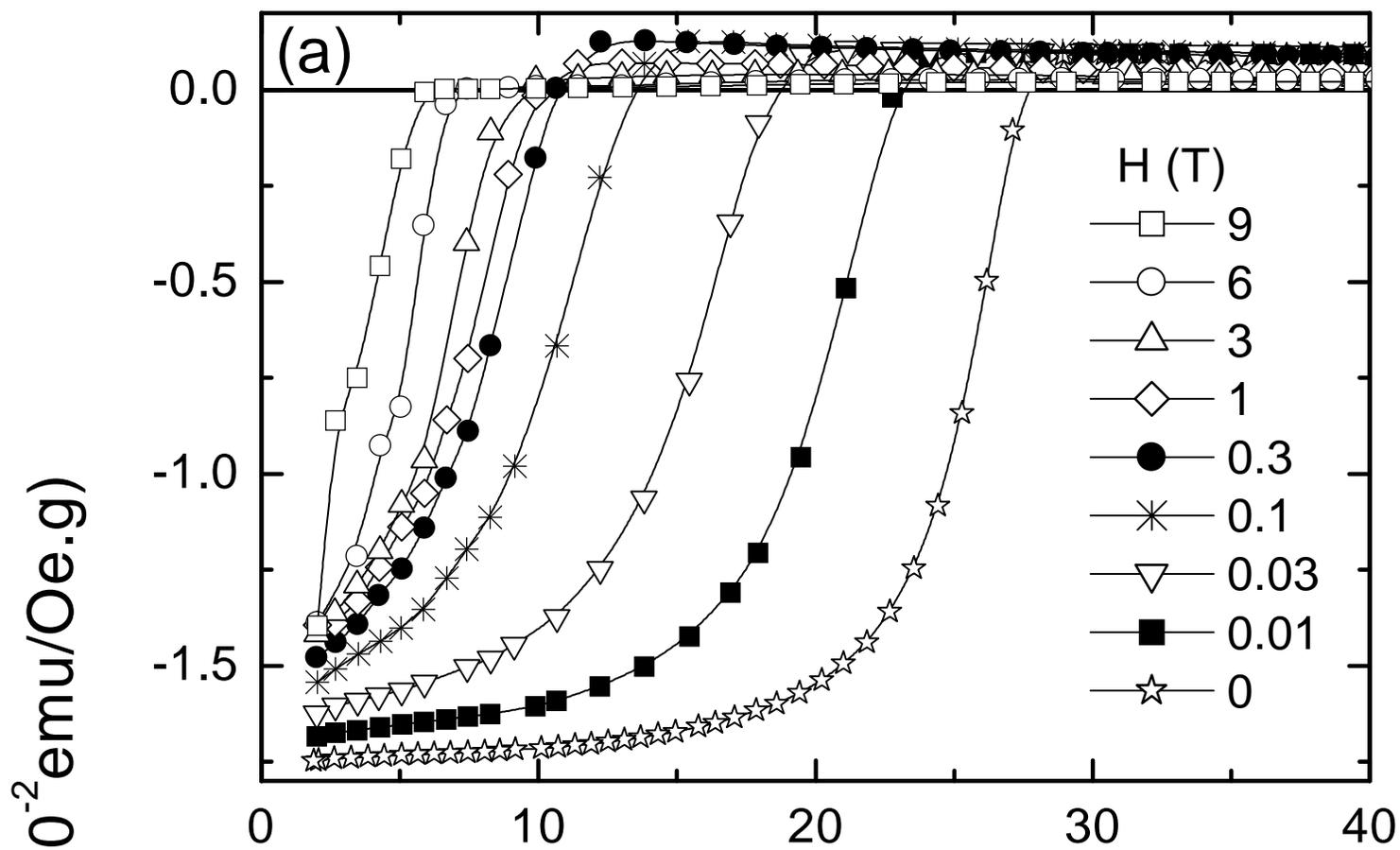
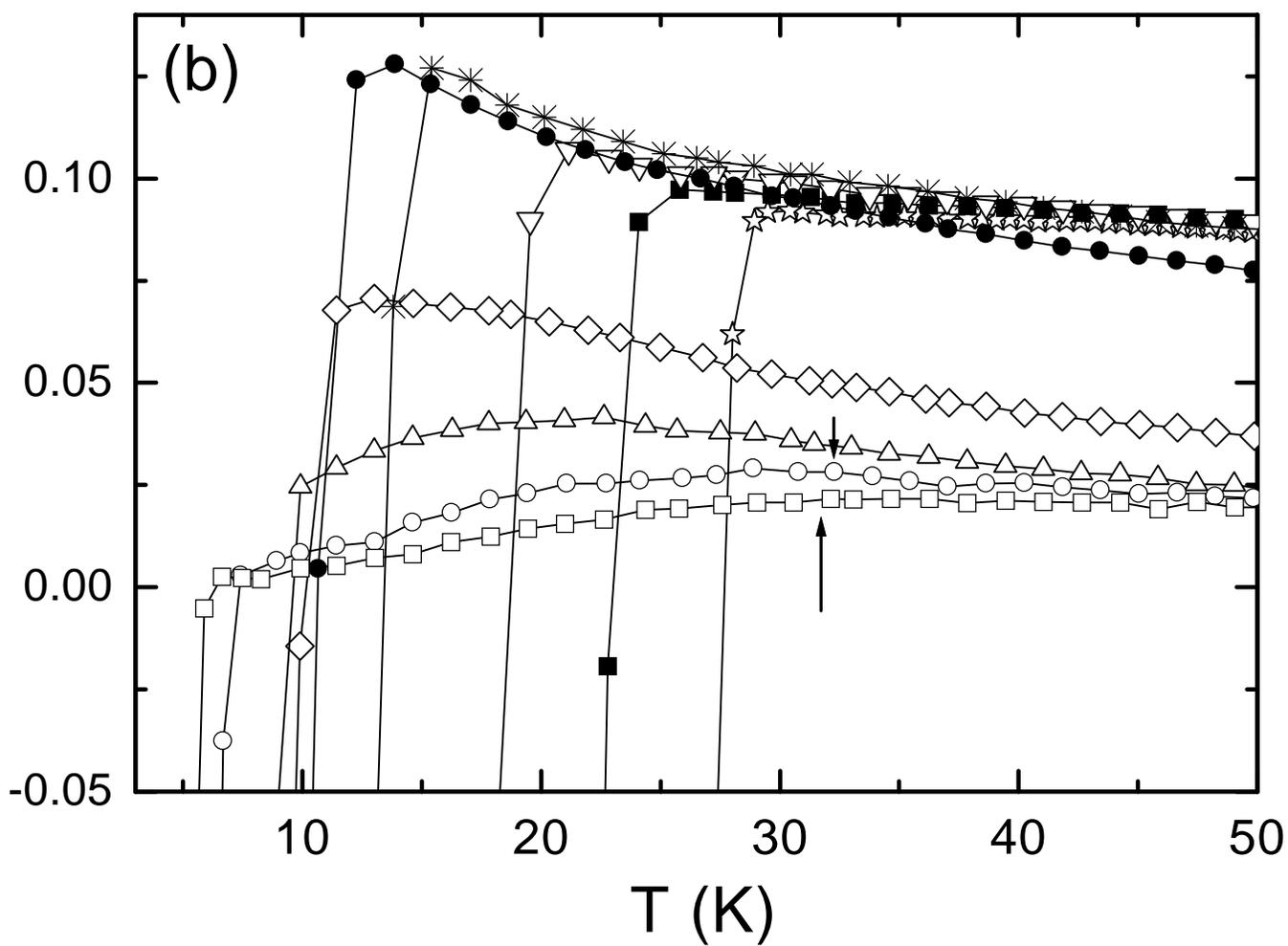
\end{document}